\documentclass[reprint, amsmath,amssymb,aps]{revtex4-1}
\usepackage{graphicx}
\usepackage{dcolumn}
\usepackage{bm}
\usepackage{float}
\usepackage{appendix}

\usepackage{bbm}
\usepackage{ulem}
\usepackage{color}
\definecolor{mygreen}{rgb}{0,0.5,0}
\definecolor{mygrey}{rgb}{0.5,0.5,0.5}
\definecolor{myred}{rgb}{0.75,0,0}
\definecolor{myblue}{rgb}{0,0,0.75}
\definecolor{mymagenta}{cmyk}{0,1,0,0.12}
\definecolor{mycyan}{cmyk}{1,0,0,0.12}
\definecolor{myorange}{rgb}{1,0.5,0}
\definecolor{myviolet}{rgb}{0.5,0.0,0.75}
\definecolor{mybrown}{rgb}{0.75,0.5,.5}

\begin{document}

\title{Quantum-enhanced Electrometer based on Microwave-dressed Rydberg Atoms}
\author{Shuhe Wu
 $^{1,2}$}
\author{Dong Zhang
 $^{1,2}$}
\author{Zhengchun Li
$^{1,2}$}
\author{Minwei Shi
$^{1,2}$}
\author{Peiyu Yang
$^{1,2}$}
\author{Jinxian Guo
$^{1,2}$}
\author{Wei Du
$^{1,2}$}
\email{wdsjtu2021@sjtu.edu.cn}
\author{Guzhi Bao
$^{1,2}$}
\email{guzhibao@sjtu.edu.cn}
\author{Weiping Zhang
 $^{1,2,3,4}$}
 \email{ wpz@sjtu.edu.cn}
\affiliation{{
$^{1}$ School of Physics and Astronomy, and Tsung-Dao Lee institute, Shanghai Jiao Tong University, Shanghai 200240, China.\\
$^{2}$ Shanghai Branch, Hefei National Laboratory, Shanghai 201315, China.\\
$^{3}$ Collaborative Innovation Center of Extreme Optics, Shanxi University, Taiyuan, Shanxi 030006, China.\\
$^{4}$ Shanghai Research Center for Quantum Sciences, Shanghai 201315, China.}}
\date{\today}

\title{Quantum-enhanced Electrometer based on Microwave-dressed Rydberg Atoms}

\date{\today}

\begin{abstract}
{Rydberg atoms have been shown remarkable performance in sensing microwave field. The sensitivity of such an electrometer based on optical readout of atomic ensemble has been demonstrated to approach the photon-shot-noise limit. However, the sensitivity can not be promoted infinitely by increasing the power of probe light due to the increased collision rates and power broadening. Compared with classical light, the use of quantum light may lead to a better sensitivity with lower number of photons. In this paper, we exploit entanglement in a microwave-dressed Rydberg electrometer to suppress the fluctuation of noise. The results show a sensitivity enhancement beating the shot noise limit in both cold and hot atom schemes. Through optimizing the transmission of optical readout, our quantum advantage can be maintained with different absorptive index of atomic vapor, which makes it possible to apply quantum light source in the absorptive electrometer.
} 

\end{abstract}
\maketitle
\section{introduction}\label{sec:introduction}

Ultrasensitive detection of electric field plays a significant role in widespread applications, including communications \cite{communication1,communication2}, remote sensing \cite{remotesensing1,remotesensing2}, and medical diagnosis \cite{medicaldiagnosis1,medicaldiagnosis2,medicaldiagnosis3}. Rydberg atom-based electrometers (RAEs) \cite{RAEs1,RAEs2}, in which the spectroscopic characterizations of atoms are engineered by the optical and electrical fields in the process of electromagnetically induced transparency (EIT) \cite{EIT1} and Autler-Townes (AT) splitting \cite{A-T2}, bring a direct International System of Units (SI)  traceable and self-calibrated measurement of electrical amplitude with broadly dynamic range. These instruments can be very sensitive to electrical fields due to the large transition electric dipole moment \cite{ dipole}, which lead to a strong atomic response to electric fields.  The strength of signal can be calculated from observing the distance between the two separate peaks caused by the electrical field \cite{RAEs1,RAEs2,1nV}.

When the electrical field is strong enough, the splitting can be distinguished at the area called AT regime.  
 However, precisely measuring the splitting will be hard when the electrical field is too small to depart from the AT regime \cite{RAEs1,1nV}. Under these circumstances, the strength of the electrical field can be monitored by observing the variation of probe transmission instead of the splitting. Recently, photon-shot-noise (PSN) in the signal readout has been regarded as a main limiting factor to the performance of RAEs \cite{PSN1,PSN2}. 
 The PSN of probe light and the slope of readout signal will jointly limit the sensitivity, which can be defined as
\begin{equation}
	\label{eq:Defsensi}
	\delta E_{\text{MW}} \equiv \frac{\sqrt{\langle\delta^2\hat{I}\rangle}}{|\partial\langle\hat{I}\rangle/\partial\,\epsilon||\partial\epsilon/ \partial\,E_{\text{MW}}|}.
\end{equation}

Here $\hat{I}$ is the intensity of the detected field, $ \epsilon$ and $E_{MW}$ represent the absorptive index and amplitude of microwave field respectively. It is clear that the PSN scales with the square root of the optical power and the slope of readout signal is proportional to the optical power, so classically optimized sensors can typically be improved by increasing the laser power. However, the laser-induced collision rates and power broadening \cite{PSN1} will lead to a decrease of signal slope which also limits the laser power. 
\begin{figure}[tbph]
\centering
\includegraphics[width=8.9cm]{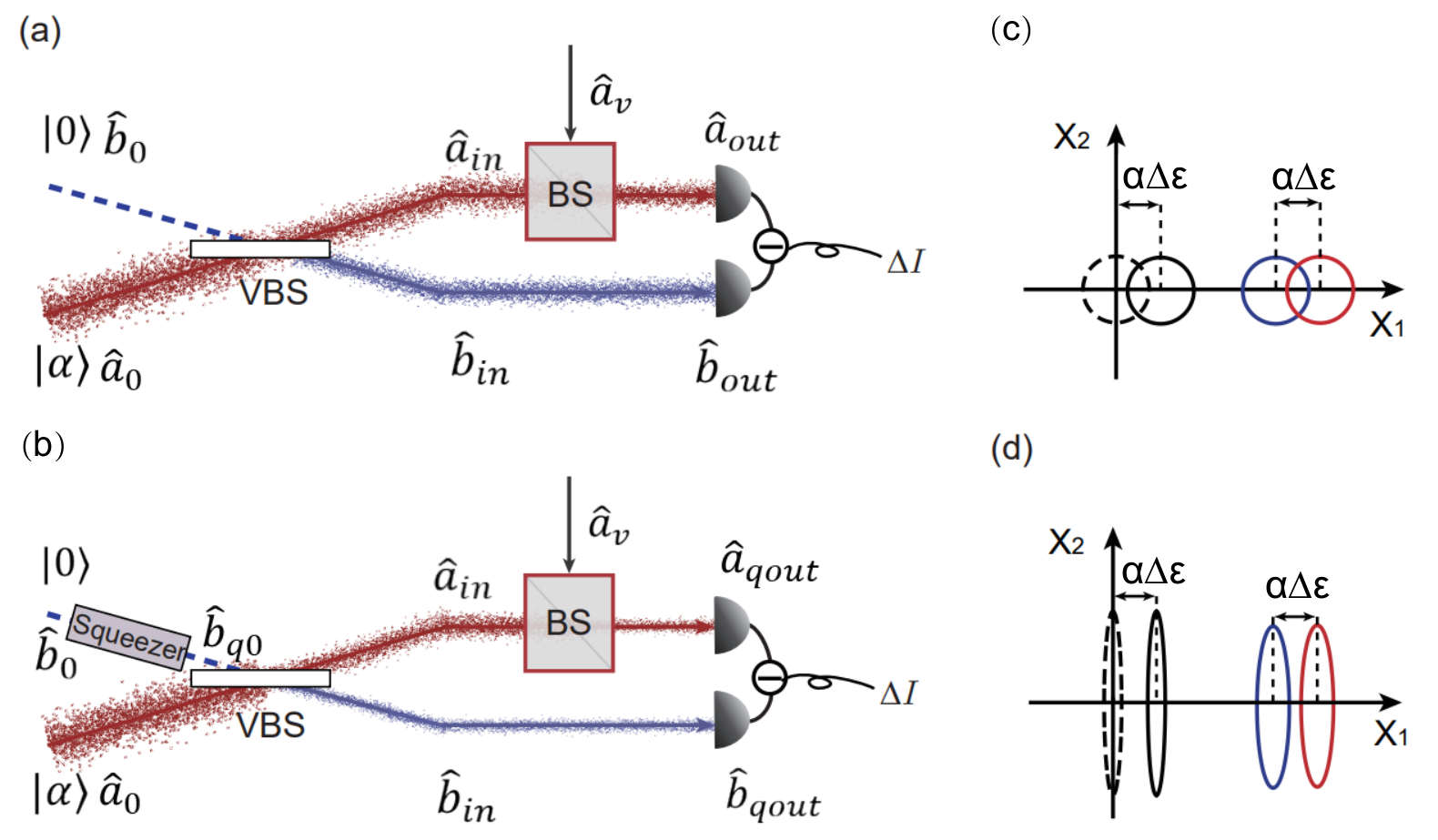}
\caption{Schematic diagram for the absorptive measurement with classical light (a) and squeezed light (b), where $|0\rangle$ represents the vacuum state and $|\alpha\rangle$ represents the coherent state, $\hat{a}_{v}$ is the vacuum field induced by the absorption measurement. VBS: variable beam splitter. BS: beam splitter. (c) and (d) show the phase-space ($X_{1}-X_{2}$) representation of the quantum state of the schematic diagram shown in (a) and (b), respectively. The red and blue circles (ellipses) represent the quantum states of $ \hat{a}_{out}(\hat{a}_{qout})$ and $ \hat{b}_{out}(\hat{b}_{qout})$, respectively. The black circle (ellipse) is the quantum state after the differential detection with $ X_{1}= X_{\hat{a}_{out}({\hat{a}_{qout}})}-X_{\hat{b}_{out}({\hat{b}_{qout}})}$ and $ X_{2}= P_{\hat{a}_{out}({\hat{a}_{qout}})}-P_{\hat{b}_{out}({\hat{b}_{qout}})}$. Here $ X_{\beta}=\beta+\beta^{\dagger}$ and $P_{\beta}=i(\beta^{\dagger}-\beta)$ with $\beta\in [\hat{a}_{out}, \hat{a}_{qout}, \hat{b}_{out}, \hat{b}_{qout}]$.} 
\label{fig:20230101}
\end{figure} 
In this scenario, squeezed light hold particular appeal for applications due to its ability to further enhance the sensitivity by reducing PSN with limited laser power, which has been shown in a number of applications \cite{Squeezedlight1,Squeezedlight2,Squeezedlight3,Squeezedlight4,Squeezedlight5}. 

In this paper, we study the absorptive measurement using the beam splitter (BS) to mimic the loss in the differential detection and theoretically analyze the characteristics of different schemes, e.g. with the coherent and squeezed input states. Then we replace the BS with microwave-dressed Rydberg atoms and study the entanglement-assisted microwave electrometer. Since quantum light is extremely fragile to loss, therefore it's necessary to find an operating point that simultaneously maintains a relatively high transmittance of probe light and slope of the observable. We demonstrate the variation of the electrical field $ E_{\text{MW}}$ affects the transmittance factor $ \epsilon$ of probe light and the slope of the observable synchronously, therefore the minimal observable is realized by operating at the optimal slope and noise through engineering the dressed microwave (MW) field and light field.
\section{sensitivity of absorptive measurement}\label{sec:smsq}
\subsection{Sensitivity with classical field}
 Before introducing the entanglement-assisted RAEs, we start by briefly calculating the sensitivity of absorptive measurement when coherent light and squeezed light are employed respectively. When coherent state is employed, the absorptive measurement can be achieved by a topology of differential detection as shown in Fig.\thinspace\ref{fig:20230101} (a). A coherent state $\hat{a}_{0}$ is splitted into $\hat{a}_{in}$ and $\hat{b}_{in}$ by the variable beam splitter (VBS) with the transmissivity of T and reflectivity of R.
$\hat{a}_{in}$ acts as the probe light propagate through a fictitious BS mimicking the absorptive signal of optical field with the transmission $ e^{-\epsilon}$ and $\hat{a}_{v}$ is the vacuum field induced by the absorptive measurement. While the beam $\hat{b}_{in}$ serves as a reference. 
The input-output relations of the coherent state scheme are given by
\begin{equation}
\begin{split}
    \hat{a}_{out}&=\sqrt{T e^{-2\epsilon}}\,\hat{a}_{0}+\sqrt{R\,e^{-2\epsilon}}\hat{b}_{0}+\sqrt{1-e^{-2\epsilon}}\,\hat{a}_{v},
\end{split}
\end{equation}
\begin{equation}
\begin{split}
    \hat{b}_{out}=\hat{b}_{in}=\sqrt{T} \,\hat{b}_{0}-\sqrt{R}\hat{a}_{0},
    \end{split}
\end{equation}
where T and R are the transmissivity and reflectivity of the VBS with the condition $T+R=1$. The observable here is defined as 
\begin{eqnarray}\label{Input-output}
\hat{I}=\hat{a}^{\dagger}_{out} \hat{a}_{out}-\hat{b}^{\dagger}_{out} \hat{b}_{out}.
\end{eqnarray}	
Then we obtain the slope and noise of the observable
\begin{equation}
\frac{\partial\langle\hat{I}\rangle}{\partial\epsilon }=-2e^{-2\epsilon}{I_{si}},\, \langle\delta^2\hat{I}\rangle=[e^{-2\epsilon}+\frac{R}{T}]{I_{si}},
\end{equation}
where, $I_{si}=T\alpha^2$ denotes the sensing intensity of absorptive measurement.
According to Eq.~(\ref{eq:Defsensi}), the sensitivity can be expressed as
\begin{equation}
\delta_{\text{c}}\epsilon =\sqrt{\frac{1}{2}+\frac{R}{2e^{-2\epsilon}T}}\frac{1}{\sqrt{e^{-2\epsilon}I_{si}}},
\end{equation}

The optimum sensitivity can be achieved when $T=1$ and $R=0$, which represents a direct detection scheme with all of the coherent state is transmitted to sense the signal. Then we get the minimum measurable 
\begin{equation}\label{clsrst}
\delta_{\text{c}} \epsilon=\frac{1}{\sqrt{2}\sqrt{e^{-2\epsilon}I_{si}}}.
\end{equation}
For practical application, the technical noise from laser also limits the sensitivity\cite{PSN1}. A differential measurement, which is commonly used in precision measurement, can be realized with $e^{-2\epsilon}T=R$ where the excess noise involved in the laser can be cancelled. 
From Eq.~(\ref{clsrst}), we find the sensitivity is restricted by the absorption induced loss and slope, the fundamental limitation is still photon shot noises (PSN) originated from the statistical distribution of photons. 

\subsection{Sensitivity with quantum field}\label{sec:smq}

\begin{figure}[H]
\centering
\includegraphics[width=8.6cm]{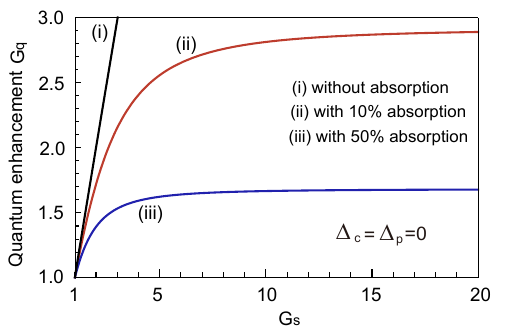}
\caption{Quantum enhancement $ G_{q}$, (i) without absorption, (ii) with 10 \% absorption and (iii) with 50\% absorption. The absorption parameters set here correspond to the optimal sensitivity of squeezed light injection with $ \Delta_{\text{c}}=\Delta_{\text{p}}=0$ (see detail in the next section). Here, we labeled the improvement factor $ G_{s}=e^{r}=\text{cosh}r+\text{sinh}r$.}
\label{fig:quantum_enhancement}
\end{figure}
Now we focus on analyzing the performance of squeezed light in absorptive measurement. As is well-known, squeezed light is difficultly generated with large photon flux and extremely sensitive to the loss in absorptive measurement. We propose a topology similar to homodyne detection, the sensing intensity can be boosted by introducing a local oscillator (LO). The optimum operation condition will be studied to realize the maximized sensitivity with precision beating the shot-noise limit (SNL). 

As shown in Fig. \thinspace\ref{fig:20230101} (b), we employ a squeezed vacuum state at the unused port of VBS to replace the vacuum.
The input-output relation of the squeezer can be expressed as 
\begin{equation}
\hat{b}_{q0}=\text{cosh} r \,\hat{b}_{0}+\text{sinh}r\,\hat{b}^{\dagger}_{0}e^{i\theta},
\end{equation}
where $\hat{b}_{0}$ is vacuum state, $\text{cosh} r$ and $\text{sinh} r$ are the gain factors satisfying $\text{cosh}^2 r-\text{sinh}^2 r=1$, $r$ is the squeezed factor and $\theta$ is phase of squeezer. The squeezed light is combined with another much stronger LO field at the VBS to boost the sensing power since it is exceedingly hard to be prepared with large photon number. For the squeezed light scheme, the input-output relationship become
\begin{equation}
\begin{split}
    \hat{a}_{qout}&=\sqrt{e^{-2\epsilon}T} \,\hat{a}_{0}+\sqrt{e^{-2\epsilon}R}\,\hat{b}_{q0}+\sqrt{1-e^{-2\epsilon}}\,\hat{a}_{v},
\end{split}
\end{equation}
\begin{equation}
\begin{split}
    \hat{b}_{qout}&=\hat{b}_{in}=\sqrt{T} \,\hat{b}_{q0}-\sqrt{ R}\,\hat{a}_{0}
    \end{split}.
\end{equation}
The mode of LO can be classically represented by $|\alpha|$ due to the much stronger power compared to squeezed vacuum state, then we have the differential intensity between the two detector
\begin{eqnarray}\label{Input-output qt}
\hat{I}_{q} &=& \hat{a}^{\dagger}_{qout} \hat{a}_{qout}-\hat{b}^{\dagger}_{qout} \hat{b}_{qout}\nonumber\\
&=&\alpha\left[ (1+e^{-2\epsilon})\sqrt{TR}\hat{X}_{b_{q0}}+\sqrt{Te^{-2\epsilon}(1-e^{-2\epsilon})}\hat{X}_{a_{v}}\right] \nonumber\\& &+\alpha^{2}(e^{-2\epsilon}T-R).
\end{eqnarray}	

Here $\hat{X}_{o}=\hat{o}+\hat{o}^{\dagger}$ with $o\in [ b_{q0}, a_{v}$] are the amplitude quadrature of the optical modes. Note that intensity term $\alpha^{2}(e^{-2\epsilon}T-R) $ can be cancelled when satisfying the condition $e^{-2\epsilon}T=R $, then we give the slope and noise
\begin{equation}
\frac{\partial\langle\hat{I}_{q}\rangle}{\partial \epsilon}=-2e^{-2\epsilon}{I_{si}},
\end{equation}
\begin{equation}
\begin{split}
\langle \delta^{2} \hat{I}_{q}\rangle&=\left[\frac{1+e^{-2\epsilon}}{e^{2r}}+1-e^{-2\epsilon}\right]e^{-2\epsilon}I_{si},
\end{split}
\end{equation}
where $ I_{si}=T\alpha^{2}=\alpha^{2}/(e^{-2\epsilon}+1)$ represent the sensing intensity. According to Eq.~(\ref{eq:Defsensi}), we give the quantum-enhanced sensitivity of absorptive measurement
\begin{equation}
 \delta_{q} \epsilon=\sqrt{\frac{1+e^{-2\epsilon}}{e^{2r}}+1-e^{-2\epsilon}}\frac{1}{\sqrt{2e^{-2\epsilon}I_{si}}},
\end{equation}

Compared to the classical light with differential detection, we find the sensitivity is improved by
 \begin{equation}\label{G_{q}}
G_{q}=\frac{\delta_{\text{c}} \epsilon}{\delta_{q} \epsilon}=\frac{\sqrt{2}}{\sqrt{\frac{1+e^{-2\epsilon}}{e^{2r}}+1-e^{-2\epsilon}}}.
\end{equation}

This improvement is same as the noise reduction originated from the squeezed light injection. The quantum-enhancement is maximal when the absorption is very small ($ \epsilon\rightarrow0 $), indicate an improvement factor of $ G_{s}=e^{r}$.

Next, we further analyze the influence of $G_{s}$ on the quantum enhancement $ G_{q}$ with the $\Delta_{\text{c}}=\Delta_{\text{p}}=0$, which is given in Fig.\thinspace\ref{fig:quantum_enhancement}. The black line (i) shows the quantum enhancement $ G_{q}$ without absorption, which is proportional to $G_{s}$. $ G_{q}$ with 10\% (ii) and 50\% (iii) absorption are shown in red (ii) and blue line (iii), respectively. The absorption parameters set here correspond to the optimal sensitivity of squeezed light injection in the case of cold and hot atoms (see detail in the next section). It is noted that with the increase of $ G_{s}$, the quantum enhancement $G_{q}$ in both cases tends to be saturated due to the existence of absorption. Based on this, we chose $r=2.5$ to measure the sensitivity. 


\section{Principle of quantum-noise limited electrometer}\label{sec:cl}
\begin{figure*}[tbph]
\centering
\includegraphics[width=18cm]{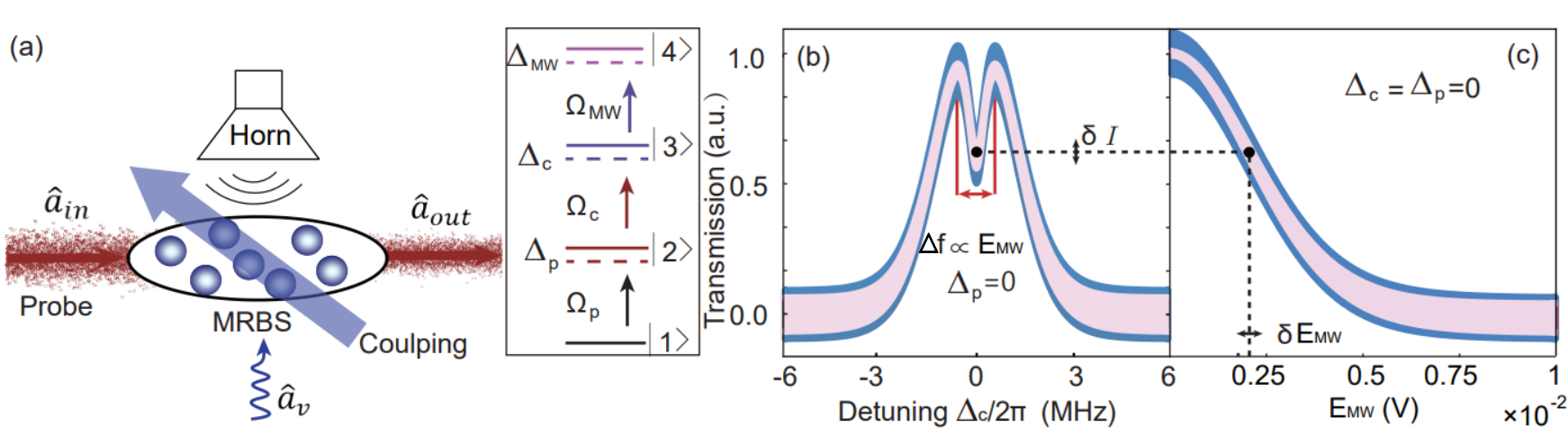}
\caption{(a) Schematic diagram for the RAEs. MRBS: Microwave-dressed Rydberg Atomic beam splitter. Horn: microwave horn, $\hat{a}_{v}$ is the vacuum field induced by the absorption measurement. Inset: Energy-level diagram of RAEs. The subscript p, c and m represent probe, coupling and MW field, $\Delta_{\text{\text{p}}}$, $\Delta_{\text{\text{\text{c}}}}$ and $\Delta_{\text{\text{MW}}}$ are their single photon detuning, $\Omega_{\text{p}}$, $\Omega_{\text{\text{c}}}$ and $\Omega_{\text{MW}}$ indicate their Rabi frequency. (b) Transmissions change with the detuning of coupling field. Here, $\Delta f$ denotes the frequency distance between the two peaks and $\Delta f \propto E_{\text{MW}}$. (c) shows the relation between transmissions and amplitude of $E_{\text{MW}}$ field when $\Delta_{\text{p}}$=$\Delta_{\text{\text{c}}}$=0. The intensity fluctuation represented by violet shade in (b) and (c) limit the sensitivity in measuring the electrical field $E_{\text{MW}}$ to shot-noise limit (blue) and squeezed quantum noise (red).}
\label{fig:Schematic diagram1}
\end{figure*}
In this theory, the RAEs are achieved by Cesium (Cs) atomic vapor with a four-level ladder structure as shown in Fig.~\ref{fig:Schematic diagram1} (a). The frequency of transition from $|1\rangle$ to $|2\rangle$  and $|2\rangle$ to $|3\rangle$ are labelled as $ \omega_{21}$ and $ \omega_{32}$ with resonance frequency of 780 nm and 480 nm, respectively. 
A weak probe light field and a strong coupling light field interact to each transition 
with frequency near atomic resonance, the quantum interference from the two excitation pathways produce a dark state which create a transparent window for the probe light field, named EIT.

A MW field with frequency $ \omega_{\text{MW}}$ resonance to the nearby Rydberg transition $|3\rangle$ to $|4\rangle$ will reduce the transmitted intensity of probe light, 
which is the physical quantity to estimate the strength of electrical field in our strategy. 
The Hamiltonian of the system after the rotating wave approximation can be written as:
\begin{equation}
\begin{array}{c}
  \hat{H}
\end{array}
=\hbar
\left[
\begin{array}{cccc}
   0 & \xi^{*}\hat{a}^{\dagger} & 0 & 0\\
   \xi\hat{a} & \Delta_{\text{p}}& \frac{\Omega^{*}_{\text{c}}}{2}& 0\\
        0 & \frac{\Omega_{\text{c}}}{2}& \Delta_{\text{c}}+\Delta_{\text{p}}& \frac{\Omega^{*}_{\text{MW}}}{2}\\
           0& 0 & \frac{\Omega_{\text{MW}}}{2} & \Delta_{\text{MW}}+\Delta_{\text{c}}+\Delta_{\text{p}}\\
\end{array}
\right],
\end{equation}
where the coupling light and MW field are treated as a classical field while the weak probe light is quantized.
The probe light are described by slowly varying quantum–mechanical operators.
$\Omega_{\text{\text{c}}}=\mu_{32}E_{\text{\text{c}}}/\hbar$ and $\Omega_{\text{\text{\text{MW}}}}=\mu_{43}E_{\text{MW}}/\hbar$ are Rabi frequency of coupling light and MW field, respectively. $ \omega_{\text{\text{c}}}, \omega_{\text{\text{p}}}$ and $\omega_{\text{\text{MW}}}$ are their frequencies. $\xi$ is the atom-probe coupling constants: $\xi=\mu_{21}\varepsilon/\hbar$, where $\mu_{ij}$ (i,j=1,2,3,4) is the transition dipole moment from state $|i\rangle$ to state $|j\rangle$ and $\varepsilon=\sqrt{\hbar\omega_{\text{\text{p}}}/2\epsilon_{0}V}$ is the electric field of a single photon. $ V$ is the quantized volume , $\hat{a}$ is the annihilation operators for probe field and  $\epsilon_{0}$ is the permittivity in vacuum. $\Delta_{\text{\text{c}}}=\omega_{32}-\omega_{\text{\text{c}}}$, $\Delta_{\text{\text{MW}}}=\omega_{43}-\omega_{\text{\text{MW}}}$, and $\Delta_{\text{p}}=\omega_{21}-\omega_{\text{\text{p}}}$ are the single photon detuning of coupling light, MW field and probe light, respectively.
The properties of the medium are described by collective, slowly-varying operators $\sigma_{\mu\nu}(z,t)=1/N_{z}\sum^{N_{z}}_{j=1}|\mu_{j}\rangle\langle\nu_{j}|e^{-i\Delta_{p}t+ik_{p}z}$ averaged over small layers denoted by their position $z$ containing number of atoms $N_{z}$. $k_{p}$ is the projection of the wavevector of probe light on the z axis.
To account for decay and dephasing, the system is described using the Heisenberg-Langevin equations:
\begin{equation}
  \frac{\partial}{\partial t} \hat{\sigma}_{\mu\nu}=\frac{i}{\hbar}[\hat{H},\hat{\sigma}_{\mu\nu}]+\hat{D}_{\mu\nu}+\hat{F}_{\mu\nu}
\end{equation}
Here, $\hat{D}_{\mu\nu}$ is the terms produced by spontaneous emission and dephasing: 
\begin{widetext}
\begin{equation}
\begin{array}{c}
  D_{\mu\nu}
\end{array}
=
\left[
\begin{array}{cccc}
 \gamma_{2}\sigma_{22}+ \gamma_{3}\sigma_{33}+\gamma_{4}\sigma_{44}  & \frac{- \gamma_{2}}{2}\sigma_{12}& \frac{- \gamma_{3}}{2}\sigma_{13}& \frac{- \gamma_{4}}{2}\sigma_{14}\\
   
\frac{- \gamma_{2}}{2}\sigma_{12} & -\gamma_{2}\sigma_{22}& -\frac{ \gamma_{2}+\gamma_{3}}{2}\sigma_{23}& -\frac{ \gamma_{2}+\gamma_{4}}{2}\sigma_{24}\\
          \frac{- \gamma_{3}}{2}\sigma_{31} & \frac{- (\gamma_{2}+\gamma_{3})}{2}\sigma_{32}& -\gamma_{3}\sigma_{33}& \frac{- (\gamma_{3}+\gamma_{4})}{2}\sigma_{34}\\
        
          \frac{- \gamma_{4}}{2}\sigma_{41}&  \frac{- (\gamma_{2}+\gamma_{4})}{2}\sigma_{42} & \frac{- (\gamma_{3}+\gamma_{4})}{2}\sigma_{43} & -\gamma_{4}\sigma_{44}\\
\end{array}
\right],
\end{equation}
\end{widetext}

Here $\gamma_{i}$ is the spontaneous decay from the states $|i\rangle$ to $|i-1\rangle$. $\hat{F}_{\mu\nu}$ is the Langevin atomic forces. 

The differential equations describing the propagation and temporal evolution of the quantum field operator is:
\begin{equation}
\begin{split}
    (\frac{\partial}{\partial t}+c\frac{\partial}{\partial z})\hat{a}(z,t)
    &=i\xi\mathcal{N}\hat{\sigma}_{12}(z,t),
\end{split}
\end{equation}
where $\mathcal{N}$ is the atomic density. The Fourier transforms of the quantum operator satisfy the following equation:
\begin{equation}
\frac{1}{k_{p}}\frac{\partial}{\partial\,z}\hat{a}(\omega)=\chi\hat{a}(w)+\hat{F}_{a},
\end{equation}
where $k_{\text{p}}$ is the wave vector of the probe, $\chi$ is the susceptibility of the medium dressed by the coupling light and MW field. $\hat{F}_{\hat{a}}=\sum_{m=2,3,4}i\xi\mathcal{N}B_{1m}\hat{F}_{1m}$. The coefficients column $[B_{1m}]$ is
\begin{equation*}
    \begin{aligned}
        &\frac{1}{S}\times
    \begin{bmatrix}
        \Omega _{\text{MW}}\Omega _{\text{MW}}^{*}+ 4\Gamma_{13} \Gamma_{14}\\
        - 2 i \Omega _\text{c}\Gamma_{14}\\
        -  \Omega _\text{c}\Omega _{\text{MW}}   
    \end{bmatrix}
    \end{aligned}.   
\end{equation*}
Here $S=\Gamma_{12}(4\Gamma_{13}\Gamma_{14}+\Omega_{\text{MW}}\Omega_{\text{MW}}^{*})+\Gamma_{14}\Omega_{\text{c}}\Omega_{\text{c}}^{*}$ with $\Gamma_{12}=i\Delta_{\text{p}}+\gamma_{2}/2, \Gamma_{13}=i(\Delta_{\text{p}}+\Delta_{\text{c}})+\gamma_{3}/2,\Gamma_{14}=i\Delta+\gamma_{4}/2$ and $\Delta=\Delta_\text{c}+\Delta _{\text{MW}}+\Delta _\text{p}$.
\begin{figure*}[t]
\centering
\includegraphics[width=18cm]{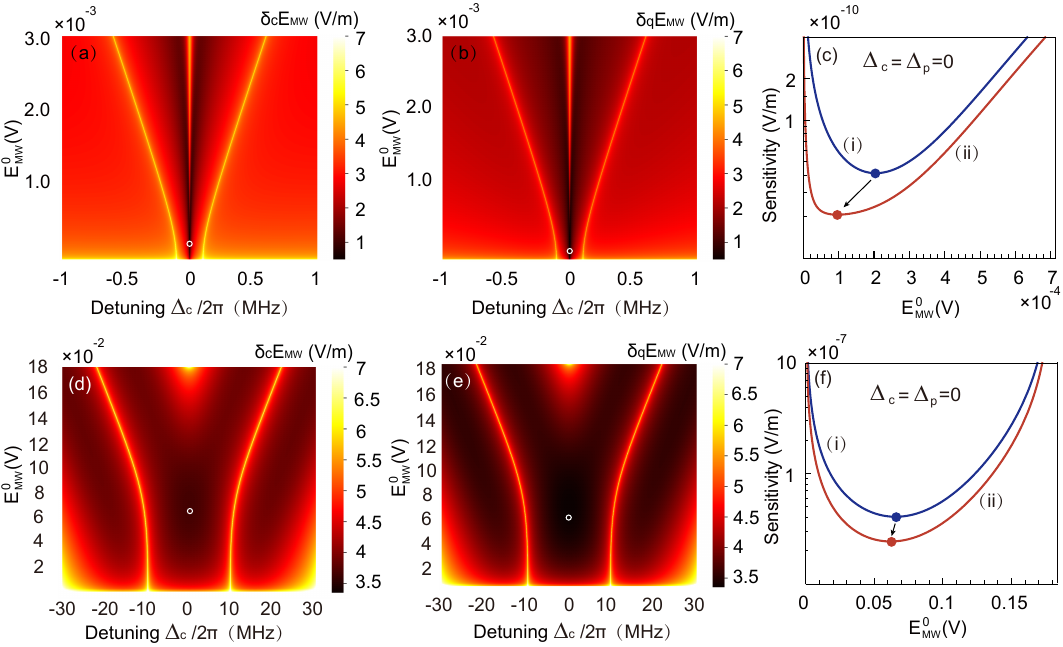}
\caption{Sensitivity with the detuning $\Delta_{\text{c}}$ and amplitude $ E_{\text{\text{MW}}}$ of MW field in the case of the cold atoms (a-c) and hot atoms (d-f), (a) and (d) coherent scheme, (b) and (e) squeezed light scheme. (c) and (f) Sensitivity change with amplitude $E_{\text{\text{MW}}}$ for $\Delta_{\text{c}}=\Delta_{\text{p}}=0$. The blue line (i) and red line (ii) represent sensitivities of classical light injection and squeezed light with $Te^{-2\epsilon}=R$, respectively. The blue and red dots denotes the optimal sensitivity of classical light and squeezed light, respectively. The black line with arrow indicates the quantum enhancement. Here, we set $\alpha=10^{7},r=2.5$.}
\label{fig:cold}
\end{figure*}
The formula of susceptibility in Rydberg atoms can be expressed as 
\begin{equation}
    \chi=\frac{i\mathcal{N}|\mu_{21}|^{2}}{\epsilon_{0}\hbar}\frac{\Gamma_{13}\Gamma_{14}+\frac{\Omega_{\text{\text{MW}}}\Omega^*_{\text{\text{MW}}}}{4}}{\Gamma_{12}[\Gamma_{13}\Gamma_{14}+\frac{\Omega_{\text{\text{MW}}}\Omega^*_{\text{\text{MW}}}}{4}]+\frac{\Omega_{\text{c}}\Omega^*_{\text{c}}}{4}\Gamma_{14}}.
\end{equation}
The formally integration of quantum operator is
\begin{equation}\label{eq:BS}
    \hat{a}_{out}=e^{-(\epsilon+i\phi)}\hat{a}_{in}+\sqrt{1-e^{-2(\epsilon+i\phi)}}\hat{a}_{\upsilon},
\end{equation}
where $\epsilon=-\text{Im}(\chi)k_{p}l$ is the absorption index, $\phi=-\text{Re}(\chi)k_{p}l$ is the dispersion index, $l$ is the length of the atomic medium. $\hat{a}_{\upsilon}(t)=\int d\tau \hat{F}_{\hat{a}}(\tau)e^{i\tau}/\sqrt{1-e^{-2(\epsilon+i\phi)t}}$. With $\gamma_{2}>>\gamma_{3}$ and $\gamma_{2}>>\gamma_{4}$ \cite{1nV}, it is easy to verify $\langle \hat{a}_{\upsilon}(t)\rangle=0$, $\langle \hat{a}_{\upsilon}(t)\hat{a}_{\upsilon}(t')\rangle=0$, $\langle \hat{a}_{\upsilon}^{\dagger}(t)\hat{a}_{\upsilon}^{\dagger}(t')\rangle=0$, $\langle \hat{a}_{\upsilon}^{\dagger}(t)\hat{a}_{\upsilon}(t')\rangle=0$ and $\langle \hat{a}_{\upsilon}(t)\hat{a}_{\upsilon}^{\dagger}(t')\rangle=\delta(t-t')$ with $ D_{1}$ and $D_{2}$ (see detail in Appendix). The input-output relation is consistent with Eq. (2) which denotes a induce of vacuum field $\hat{a}_{\nu}$ in the loss measurement. 

From the above, we note that the transmission of probe light is a function of MW field $E_{\text{\text{MW}}}$. Obviously, the intensity fluctuation and slope simultaneously affect the sensitivity for measuring microwave. At the position with a large slope, we can sensitively measure the MW field by observing the intensity of transmitted probe light, as shown in the  Fig.\thinspace\ref{fig:Schematic diagram1} (b) and (c). 

\section{Sensitivity Optimization}\label{sec:optimal}

In this chapter, we will analyze the optimal operating point of RAEs in classical light and squeezed light schemes respectively, and give their best sensitivity. As we have described above, the maximum quantum enhancement can be reached when $\epsilon\rightarrow0$. However, the slope $\partial\langle\hat{I}\rangle\,/\partial\,E_{\text{\text{MW}}}$ at this point approach zero, therefore leading to a poor sensitivity. In order to sensitively measure the weak MW field, it is necessary to find an operating point with a large slope in the classical scheme, since the noise and slope jointly determine the sensitivity as shown in Fig.~\ref{fig:Schematic diagram1}(c). We employ a dressed MW field $ E_{\text{\text{MW}}}^{0}$ whose frequency resonant with Rydberg transitions. The dressed MW field cause the splitting of transmitted peak, therefore engineering the transmission and slope of the probe. We focus on the ability of the system in sensing a very weak MW field $\delta E_{\text{\text{MW}}}$ with a dressed MW field $ E_{\text{\text{MW}}}^{0}$, here we define a MW field $ E_{\text{\text{MW}}}=E_{\text{\text{MW}}}^{0}+\delta E_{\text{\text{MW}}}$. The sensitivity in measuring $\delta E_{\text{\text{MW}}}$ field depends on $\Delta_{\text{c}}$ and $E_{\text{MW}}^{0}$. Since
the $E_{\text{MW}}^{0}$ determines the distance between the two peak $\Delta f$ and $\Delta_{\text{c}}$ leads to an asymmetry of the splitting peaks.



In Fig.\thinspace\ref{fig:cold} (a) and (b), we compare the performance in measuring MW field of cold RAEs by coherent light and squeezed light injected schemes, respectively. The sensitivity is plotted as a function of the detuning of coupling light $\Delta_{\text{c}}$ and strength of dressed MW field $ E_{\text{MW}}^{0}$, the optimal sensitivity is labelled with a white circle where $\Delta_{\text{c}}=0 $, $ \Delta_{\text{p}}=0$, and a small dressed MW field is applied. In order to reveal the difference between quantum and classical strategies more clearly, their sensitivity evolve with the dressed MW field when $\Delta_{\text{c}}=0$ is shown in Fig.\thinspace\ref{fig:cold} (c).
Employing the squeezing light can improve the sensitivity compared with the classical light with different $ E_{\text{MW}}$, as shown by the red line (ii) and the blue line (i). 
In the case of squeezed light, the optimum sensitivity is  2.1 $ \times 10^{-11}  \text{V/m}$ with a dressed $ E_{\text{MW}}^{0}=1 \times 10^{-4} \text{V}$, which has $ G_{q}=3$ compared with classical light according to Eq.~(\ref{G_{q}}).
While the optimal sensitivity of classical light case is $4.1 \times 10^{-11} \text{V/m}$ at $E_{\text{MW}}^{0}= 2 \times 10^{-4} \text{V} $. The noise of the squeezed light is extremely sensitive to the loss, which leads to the inconsistency of the optimal sensitivity points compared with the classical light. 
A suitable photon number $ \alpha= 10^{7}$ \cite{photonnumber} is chosen here to optimize the sensitivity of RAES by balancing the collision rates and power broadening from increasing the laser power.

Since the thermal motion of atoms are inevitable, Doppler broadening should be considered when the system runs at room temperature. Next, We discuss the influence from Doppler effect and give the optimal sensitivity in the case of hot atomic vapor. The atoms in the vapor satisfy Maxwell-Boltzmann distribution of
velocities, 
\begin{equation}
  f(v)=\sqrt{\frac{M}{2\pi k_{B}T}} e^{\frac{-Mv^2}{2k_{B}T}} 
\end{equation}
where $ M$ is the mass of atom, $k_{B}$ is the Boltzmann constant, $ v $ is the velocities and $T$ is temperature. The atoms move with different velocities leading to a revision of the detuning,
\begin{equation}
\Delta^{'} _{\text{\text{c}}}=\Delta _{\text{c}}-\kappa_{\text{c}}f(v),  
\\
\Delta^{'} _{\text{p}}=\Delta _{\text{p}}+\kappa_{\text{p}}f(v)
\end{equation}
Here, $\kappa_{\text{c}}$ and $\kappa_{\text{p}}$ are the wave number of probe and coupling light, respectively.

Fig.\thinspace\ref{fig:cold} (d-f) give the relationship between the detuning $\Delta_{\text{\text{c}}}$, amplitude $ E_{\text{MW}}$ of MW field and sensitivity of classical light (d) and squeezed light (e) injection in the case of hot atoms. For hot atoms, the trend of the sensitivity is consistent with that of cold atoms. Although the number of hot atoms increases by three orders of magnitude relative to the number of cold atoms, due to the  distribution of velocities broaden the transmission spectrum leads to the smaller slope, and larger absorption, which result in a smaller signal and a decrease in quantum enhancement, respectively. The overall variation leads to a worse sensitivity compared with cold atoms.
We fixed the detuning $\Delta_{\text{\text{c}}}=0$, and scanning the $E_{\text{MW}}^{0}$, as shown in Fig.\thinspace\ref{fig:cold} (f). 
The optimal sensitivity of the RAEs with the classical and squeezed light injection is  $4\times 10^{-8} \text{V/m}$ and $2.4\times 10^{-8} \text{V/m}$, respectively, at about $E_{\text{MW}}^{0}=6.5 \times 10^{-2} \text{V}$.

\section{Conclusion}\label{sec:Conclusion}
In summary, we have investigated a absorptive measurement scheme that beat the limitation of PSN by employing squeezed light. The quantum enhancement increases linearly with gain when there is no absorption, and tends to saturate with gain when absorption exists. Loss is the main limitation for the current strategy to fully utilize the injected quantum resource. Furthermore, we have studied the entanglement-assisted microwave electrometer in the Rydberg atomic by employing squeezed light with the absorptive measurement scheme. The noise squeezing makes it possible to break the bottleneck that the limited power of the sensing field restricts the sensitivity due to the laser-induced collision rates and power broadening in the atomic system. Our theoretical analysis shows the quantum advantage in our strategy is possibly maintained in absorptive sensor at large range of measuring MW field by optimally choosing the operating parameters of VBS, optical field, and atomic vapor. Squeezed-light-assisted RAEs outperform the classical scenario when the atomic vapor is operated at both cold and hot.  We notice that RAEs also induce the change of phase of probe field. The MW field sensing can also be achieved by quantum enhanced phase measurement.
Our research paves the way for the promising outlook of quantum light source in the field of absorptive measurement for future.

\section*{Fundings}
We acknowledge financial support from the Innovation Program for Quantum Science and Technology (2021ZD0303200), the National Natural Science Foundation of China (grant nos. 12234014, 12204304, 11904227, and 11654005), the Shanghai Municipal Science and Technology Major Project (grant no. 2019SHZDZX01), the Fellowship of China Postdoctoral Science Foundation (grant nos. 2020TQ0193, 2021M702146, 2021M702150, 2021M702147, and 2022T150413), the Sailing Program of the Science and Technology Commission of Shanghai Municipality (19YF1421800), the Fundamental Research Funds for the Central Universities, and the National Key Research and Development Program of China (grant no. 2016YFA0302001). W.Z. acknowledges additional support from the Shanghai Talent Program.

\section*{Disclosures}

The authors declare that there are no conflicts of interest related to this article.

\section{Appendix} 
\subsection{Steady state} \label{appendix A}

In zeroth-order perturbation expansion, in which $\hat{a}$ go to zero, the Heisenberg-Langevin equations for $\hat{\sigma}_{11}, \hat{\sigma}_{22}, \hat{\sigma}_{33}, \hat{\sigma}_{44}, \hat{\sigma}_{23}, \hat{\sigma}_{32}, \hat{\sigma}_{24}, \hat{\sigma}_{42}, \hat{\sigma}_{34}, \hat{\sigma}_{43}$ atomic operators are decoupled. The mean values of these operators are required for the next order solution. We assume the coupling light and MW field to propagate without depletion, as we verified numerically. Then the subset of equations for the mean value variables $ \langle\hat{\sigma}_{11}\rangle, \langle\hat{\sigma}_{22}\rangle,\langle\hat{\sigma}_{33}\rangle,\langle\hat{\sigma}_{44}\rangle, \langle\hat{\sigma}_{23}\rangle, \langle\hat{\sigma}_{32}\rangle, \langle\hat{\sigma}_{24}\rangle, \langle\hat{\sigma}_{42}\rangle, \langle\hat{\sigma}_{34}\rangle, \langle\hat{\sigma}_{43}\rangle$ to be solved at the steady state is written in matricial form as following:
\begin{equation}
\label{eq:appendix A1}
 ({[I]_{9\times9}\frac{\partial}{\partial t}-[M_{0}]})[\Sigma_{0}]=[S_{0}], 
\end{equation}

where $ [I]_{9\times9}$ is the 9 $\times$ 9 identity matrix.
\begin{widetext}
\begin{equation}
\begin{array}{c}
  {M_{0}}
\end{array}
=
\left[
\begin{array}{ccccccccc}
  -\gamma_{4} & \gamma_{2}-\gamma_{4} & \gamma_{3}-\gamma_{4} & 0& 0& 0& 0& 0& 0\\
  0 & -\gamma_{2}& 0& \frac{-i\Omega_{\text{c}}}{2}& \frac{i\Omega_{\text{c}}}{2}& 0& 0& 0& 0\\
 0 & 0& -\gamma_{3}& \frac{i\Omega_{\text{c}}}{2}& \frac{-i\Omega_{\text{c}}}{2}& 0& 0& \frac{\Omega_{\text{MW}}}{2}& \frac{-\Omega_{\text{MW}}}{2}\\
  0 &  \frac{-i\Omega_{\text{c}}}{2}& \frac{i\Omega_{\text{c}}}{2}& \frac{i2\Delta_{\text{c}}-\gamma_{3}-\gamma_{2}}{2}&0 & \frac{-i\Omega_{\text{MW}}}{2}& 0&0 & 0\\
  0 &  \frac{i\Omega_{\text{c}}}{2}& \frac{-i\Omega_{\text{c}}}{2}& 0&\frac{ -i2\Delta_{\text{c}}-\gamma_{3}-\gamma_{2}}{2} & 0&\frac{i\Omega_{\text{MW}}}{2}&0 & 0\\
   0 & 0& 0& \frac{-i\Omega_{\text{MW}}}{2}& 0&\frac{-\beta_{1}}{2}&0 &\frac{i\Omega_{\text{c}}}{2}&0\\
      0 & 0& 0& 0& \frac{i\Omega_{\text{MW}}}{2}&0&\frac{-\beta_{2}}{2} &0&\frac{-i\Omega_{\text{c}}}{2}\\
      \frac{-i\Omega_{\text{MW}}}{2} &   \frac{-i\Omega_{\text{MW}}}{2}& -i\Omega_{\text{MW}}& 0& 0& \frac{i\Omega_{\text{c}}}{2} &0&\frac{ -\gamma_{3}-\gamma_{4}-i2\Delta_{\text{MW}}}{2}&0\\
      \frac{i\Omega_{\text{MW}}}{2} &   \frac{i\Omega_{\text{MW}}}{2}&  i\Omega_{\text{MW}}& 0& 0& 0 &\frac{-i\Omega_{\text{c}}}{2}&0&\frac{ -\gamma_{3}-\gamma_{4}+i2\Delta_{\text{MW}}}{2}\\
\end{array}
\right],
\end{equation}
\end{widetext}
\begin{equation}
\begin{array}{c}
  {[\hat{\Sigma}_{0}]}
\end{array}
=
\left[
\begin{array}{c}
\hat{\sigma}_{11} \\\hat{\sigma}_{22} \\ \hat{\sigma}_{33}\\ \hat{\sigma}_{23}
 \\\hat{\sigma}_{32} \\ \hat{\sigma}_{24} \\ \hat{\sigma}_{42} \\ \hat{\sigma}_{34} \\ \hat{\sigma}_{43}
\end{array}
\right],
\begin{array}{c}
  {[S_{0}]}
\end{array}
=
\left[
\begin{array}{c}
\gamma_{4}\\0 \\ 0\\ 0 \\0 \\ 0 \\ 0\\ \frac{i\Omega_{\text{MW}}}{2} \\ \frac{-i\Omega_{\text{MW}}}{2}
\end{array}
\right].
\end{equation}
where $\beta_{1}= \gamma_{2}+\gamma_{4}+i2\Delta_{c}+i2\Delta_{\text{MW}}$ and $\beta_{2}= \gamma_{2}+\gamma_{4}-i2\Delta_{c}-i2\Delta_{\text{MW}}$. The steady state solution of Eq.~(\ref{eq:appendix A1}) is
\begin{equation}
[\langle\Sigma_{0}\rangle]=[M_{0}]^{-1}[S_{0}],
\end{equation}
\subsection{Atomic Heisenberg-Langevin equations} \label{appendix B}
  
The first order solution for the three coherences $\sigma_{12},  \sigma_{13},  \sigma_{14}$ is determined by the following matricial equation
\begin{equation}
\label{eq:appendix B1}
({[I]_{3\times3}\frac{\partial}{\partial t}-[M_{1}]})[\hat{\Sigma}_{1}]=[S_{1}]\hat{a}+[\hat{F}_{1}],
\end{equation}
with 
\begin{equation}
\begin{array}{c}
  {M_{1}}
\end{array}
=
\left[
\begin{array}{ccc}
  \frac{-\gamma_{2}}{2}-i\Delta_{\text{p}}& \frac{-i\Omega_{\text{c}}}{2}& 0\\
  \frac{-i\Omega_{\text{c}}}{2}& \frac{-\gamma_{3}}{2}-i(\Delta_{\text{p}}+\Delta_{\text{c}})& \frac{-i\Omega_{\text{MW}}}{2}\\
 0& \frac{-i\Omega_{\text{MW}}}{2}&  \frac{-\gamma_{4}}{2}-i\Delta\\
\end{array}
\right],
\end{equation}

\begin{equation}
\begin{array}{c}
  {[\hat{\Sigma}_{1}]}
\end{array}
=
\left[
\begin{array}{c}
\hat{\sigma}_{12} \\\hat{\sigma}_{13} \\ \hat{\sigma}_{14}
\end{array}
\right],
\begin{array}{c}
  {[S_{1}]}
\end{array}
=\xi
\left[
\begin{array}{c}
i\langle\hat{\sigma}_{22}-\hat{\sigma}_{11}\rangle\\
i\langle\hat{\sigma}_{23}\rangle \\
i\langle\hat{\sigma}_{24}\rangle
\end{array}
\right],
\\
\newline
\\
\begin{array}{c}
  {[\hat{F}_{1}]}
\end{array}
=
\left[
\begin{array}{c}
\hat{F}_{12} \\\hat{F}_{13} \\ \hat{F}_{14}
\end{array}
\right].
\end{equation}

Here, $ [I]_{3\times3}$ is the 3 $\times$ 3 identity matrix. The annihilation operators being denoted $ \hat{a}$, and $\Delta=\Delta_{\text{\text{p}}}+\Delta_{\text{c}}+\Delta_{\text{MW}}$. The Langevin atomic forces $ [\hat{F}]$ are characterized by their
diffusion coefficients matrix $ [D_{1}]+[D_{2}]$, defined as
\begin{equation}
    [D_{1}]2\delta(t-t')\delta(z-z')=\langle[\hat{F}_{1(z,t)}][\hat{F}_{1(z,t')}^{\dagger}]\rangle
\end{equation}
\begin{equation}
    [D_{2}]2\delta(t-t')\delta(z-z')=\langle[\hat{F}_{1(z,t')}^{\dagger}][\hat{F}_{1(z,t)}]\rangle
\end{equation}
Langevin diffusion coefficients for operators can be calculated
using the generalized Einstein relation\cite{Langevin}. The
$[D_{1}]$ and $[D_{2}]$ diffusion matrices are given by
 \begin{equation}
\begin{array}{c}
  {D_{1}}
\end{array}
=
\left[
\begin{array}{ccc}
  \gamma_{2}& 0& 0\\
  0& \gamma_{3}&0\\
 0& 0&  \gamma_{4}\\
\end{array}
\right],
\begin{array}{c}
  {D_{2}}
\end{array}
=
\left[
\begin{array}{ccc}
 0& 0& 0\\
  0& 0&0\\
 0& 0& 0\\
\end{array}
\right].
\end{equation}
By linearizing Eq.~(\ref{eq:appendix B1}) we derive for the mean values
 \begin{equation}
     [\langle \hat{\Sigma}_{1} \rangle]=-[M_{1}]^{-1}[S_{1}] [\langle \hat{a}\rangle]
 \end{equation}
 and for the Fourier-transformed quantum fluctuations
 \begin{equation}
   [\delta\hat{\Sigma}_{1}]=-([M_{1}]+i\omega[I]_{3\times3})^{-1}[S_{1}][\delta\hat{a}]-([M_{1}]+i\omega[I]_{3\times3})^{-1}[\hat{F}_{1}]
 \end{equation}
 Here, $\omega$ is the analysis frequency.

\end{document}